\documentclass[
prl
,reprint%
,twocolumn%
,secnumarabic%
,floatfix%
,superscriptaddress
,showpacs%s
,amssymb, amsmath, nobibnotes, aps]{revtex4-2}

\usepackage[latin1]{inputenc}
\usepackage{epsfig}
\usepackage[usenames,dvipsnames]{color}
\usepackage{comment}
\usepackage{appendix}
\usepackage{esint}
\usepackage{comment}

\newcommand{\psirt}{\psi(\mathbf r,t)}

\usepackage{braket}
\usepackage{booktabs}
\usepackage{pdfpages}
   \makeatletter
    \AtBeginDocument{\let\LS@rot\@undefined}
    \makeatother
\usepackage{pgffor}

\usepackage{xr-hyper}
\makeatletter
\newcommand*{\addFileDependency}[1]{
  \typeout{(#1)}
  \@addtofilelist{#1}
  \IfFileExists{#1}{}{\typeout{No file #1.}}
}
\makeatother

\newcommand*{\myexternaldocument}[1]{
    \externaldocument{#1}
    \addFileDependency{#1.tex}
    \addFileDependency{#1.aux}
}
\myexternaldocument{./supplementary}
\usepackage{hyperref}
\hypersetup{
    colorlinks,
    linkcolor={blue},
    filecolor={red},
    citecolor={blue},
    urlcolor={blue},
    breaklinks
}

\begin{document}

\title{Emergent Universal Drag Law in a Model of Superflow}

\author{M.T.M. Christenhusz}
\email{m.christenhusz@uq.edu.au}
\affiliation{Australian Research Council Centre of Excellence for Engineered Quantum Systems, School of Mathematics and Physics, University of Queensland, St. Lucia, QLD 4072, Australia.}
\author{A. Safavi-Naini}
\affiliation{Institute of Physics, University of Amsterdam, Science Park 904, 1098 XH Amsterdam, the Netherlands.}
\author{H. Rubinsztein-Dunlop}
\affiliation{Australian Research Council Centre of Excellence for Engineered Quantum Systems, School of Mathematics and Physics, University of Queensland, St. Lucia, QLD 4072, Australia.}
\author{T.W. Neely}
\affiliation{Australian Research Council Centre of Excellence for Engineered Quantum Systems, School of Mathematics and Physics, University of Queensland, St. Lucia, QLD 4072, Australia.}
\author{M.T. Reeves}
\affiliation{Australian Research Council Centre of Excellence in Future Low-Energy Electronics Technologies, School of Mathematics and Physics, University of Queensland, St Lucia, QLD 4072, Australia.}
\date{\today}

\begin{abstract}
Despite the fundamentally different dissipation mechanisms, many laws and phenomena of classical turbulence equivalently manifest in quantum turbulence. The Reynolds law of dynamical similarity states that two objects of same geometry across different length scales are hydrodynamically equivalent under the same Reynolds number, leading to a universal drag coefficient law. In this work we confirm the existence of a universal drag law in a superfluid wake, facilitated by the nucleation of quantized vortices. We numerically study superfluid flow across a range of Reynolds numbers for the paradigmatic classical hard-wall and the Gaussian obstacle, popular in experimental quantum hydrodynamics. In addition, we provide a feasible method for measuring superfluid drag forces in an experimental environment using control volumes.
\end{abstract}

\maketitle
%\section{Introduction}

The principle of dynamical similarity is a cornerstone of classical fluid dynamics that allows problems to be cast in terms of dimensionless parameters. This principle is not just a useful heuristic of dimensional analysis but is intimately tied to the scale invariance of the governing equations, revealing underlying universality of different flow phenomena. Two physical situations are guaranteed to be identical under appropriate scaling of space and time only if all relevant control parameters are the same~\cite{batchelor_1967_Ch4_7,Frisch1995Turbulence}.
The Reynolds number, $\mathrm{Re}$, is perhaps the most ubiquitous control parameter, governing the transition from laminar (smooth) to turbulent (vorticity dominated) flows. The Reynolds number for flow velocity $u$, characteristic length $D$, and kinematic viscosity $\nu$, is given by  $\mathrm{Re} = uD/\nu$, and measures the ratio of inertia to viscosity, characterising the degree of turbulent motion.

Superfluid turbulence displays significant similarities to its classical counterpart. These include the persistence of the celebrated Kolmogorov law~\cite{Nore1997Kolmogorov,maurerj.LocalInvestigationSuperfluid1998,Araki2002Kolmogorov,Kobayashi2005Kolmogorov,salortTurbulentVelocitySpectra2010,PhysRevB.86.104501}, the same value of the Kolmogorov constant, and the existence of the dissipation anomaly ~\cite{kolmogorov1941dissipationanomaly,galantucci2023dissipation,makinenRotatingQuantumWave2023,dissipationAnomalyNote}. Further hydrodynamical similarities are observed in decaying and grid turbulence~\cite{stalpDecayGridTurbulence1999}, and in the velocity probability density function~\cite{mantiaQuantumClassicalTurbulence2014} in He-II systems. Moreover, in a superfluid the vortex dynamics are governed by the Kirchoff/Biot-Savart equations, which remain invariant under the same scaling transform as the Navier-Stokes equations~\cite{Scharz1985BiotSavart,Fetter1966}. 
Beyond these examples relevant to high Reynolds numbers, the similarities extend further still; for example, the von K\'arm\'an street --- a moderate Reynolds number phenomenon --- emerges in the wake of a bluff body in a pure superfluid~\cite{Sasaki2010,kwon2016expvortexstreet}. These similarities suggest that an appropriate Reynolds number should emerge in the pure superfluid. However, superfluids differ from ordinary fluids in three major respects: i) their viscosity $\nu=0$ identically; ii) vorticity may only be introduced with discrete circulation $\Gamma = 2\pi \hbar /m$, and iii) vorticity may only appear above a critical velocity $u_c$. Thus any Reynolds number should account for the close connection between vortex shedding and effective viscosity of the superfluid once the flow velocity surpasses a critical velocity $u_c$. As noted by Onsager, circulation quantum has the same dimensions as $\nu$ suggesting $\mathrm{Re}_S \sim uD/\kappa$, with $\kappa = h/m$~\cite{Onsager}. While the utility of this definition has been demonstrated in Ref. ~\cite{finneIntrinsicVelocityindependentCriterion2003}, the absence of $u_c$ makes it challenging to show universal behaviour particularly at low $\textrm{Re}_S$. In Ref.~\cite{Reeves2015} it was shown that when accounted for the critical velocity, the ``superfluid Reynolds number" $\mathrm{Re}_S = (u-u_c)D/\kappa$~\cite{SFReynolds}, revealed similarity in the vortex shedding frequency behind a Gaussian-shaped obstacle and  was subsequently utilized to characterize intermittent turbulence in superfluid He-II~\cite{Schoepe2015HeIntermittency}. However, to establish dynamical similarity through this superfluid Reynolds number we must first answer fundamental open questions: i) are \emph{all} relevant dimensionless quantities expressible as a universal function of $\textrm{Re}_S$? and ii) is the universality robust to the system details embedded within $u_c$?~\cite{Takeuchi2024}. 

In this Letter we address these fundamental open questions by numerically investigating the drag force, $F_d$, experienced by different obstacles within a 2D superflow within the Gross-Pitaevskii model. We identify different regimes of drag associated with differing regimes of vortex shedding and investigate the drag coefficient $C_d$. Next, we study the flow across a range of Reynolds numbers for several obstacles. We account for the dependence of $u_c$ on the obstacle shape to extract a universal relation for $C_d$. Finally, we introduce an experimental method for measuring $F_d$ in current platforms which will allow for the confirmation of the universal relations obtained here.

\begin{figure*}[!t]
\includegraphics[width=\textwidth,trim= 0cm 0cm 0cm 0cm, clip=true]{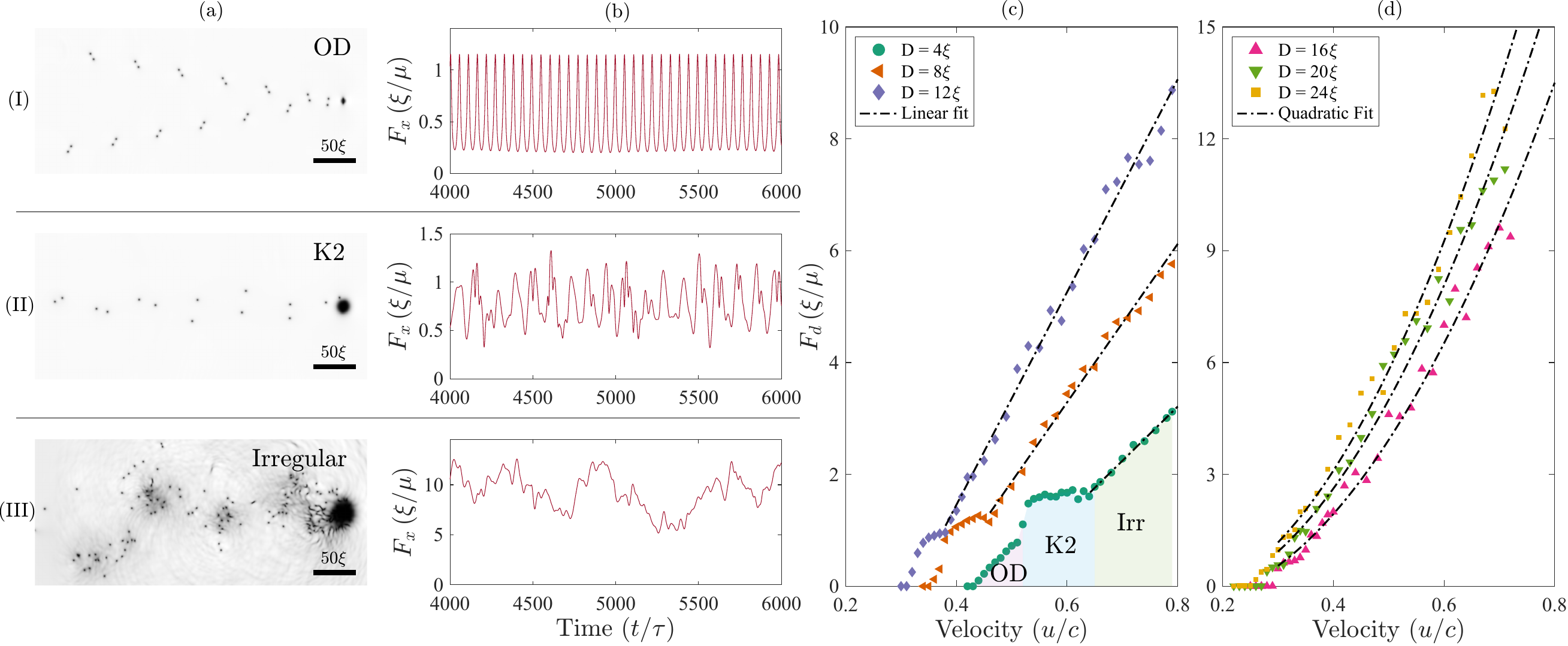}
\caption{(a) The three vortex shedding regimes (OD, K2, Irregular) in superfluid wakes with (b) corresponding longitudinal force time series for a Gaussian obstacle of width $D = 4\xi$ (I), $D = 12\xi$ (II) and $D = 24\xi$ (III). Panel (c) depicts the mean of the steady state force-time series for obstacle sizes, $4\xi$, $8\xi$ and $12\xi$. For these small obstacle sizes, the different shedding regimes can be observed and drag scales linearly (dashed fits). (d) Upon increasing the obstacle size and Reynolds number, quadratic drag scaling emerges (dashed guidelines) for obstacle sizes $16\xi$, $20\xi$ and $24\xi$.}

\label{fig:fig1}
\end{figure*}

We consider the flow of a 2D superfluid past an obstacle, modelling by the Gross-Pitaevskii equation (GPE). In the frame in which the obstacle is at rest inside a moving superfluid at velocity $\mathbf u$, the equation of motion is:

\begin{equation}
\label{eq:GPEMovingFrame}
    i\hbar \frac{\partial\psi(\mathbf{r},t)}{\partial t} = \left( \mathcal{L} - \mathbf{u}\cdot\mathbf{p}-\mu \right)\psi(\mathbf{r},t),
\end{equation}
where $\mu$ is the chemical potential, $\mathbf{p} = -i\hbar\nabla$, and
\begin{equation}
    \mathcal{L}\psirt \equiv \left[  -\frac{\hbar^2}{2m} \nabla^2 + V(\mathbf r) + g_2\vert \psirt  \vert^2  \right]\psirt.
    \label{eq:GPE}
\end{equation}
Here $V(\mathbf r)$ is functional form of the obstacle, and $g_2$ is an effective 2D interaction strength. We work in natural units of the healing length $\xi$, and time $\tau$, where we set $\mu = 1$.

The drag force $F_d$ may be calculated from the Ehrenfest relation, $F_x(t) = -\int d^2\mathbf{r}\; \psi^*(\mathbf{r},t) (\partial_x V(\mathbf{r})) \psi(\mathbf{r},t)$, which may then be averaged over time once the system reaches a steady state: $F_d = \langle F_x(t) \rangle$.  To achieve steady state wakes we implement the fringe method introduced in Ref.~\cite{Reeves2015}, which adds a damping layer $\gamma(\mathbf{r})$ at the edge of the domain, and deletes vortices within the fringe region using a phase imprinting technique~\cite{PhysRevLett.112.145301}.  This allows recycling of the flow on periodic boundary conditions, and facilitates the long-time simulations required to accurately resolve the drag force.  

This modifies Eq.~(\ref{eq:GPEMovingFrame}) to
\begin{equation}
    i\hbar \frac{\partial\psi(\mathbf{r},t)}{\partial t} = \left( \mathcal{L} - \mathbf{u}\cdot\mathbf{p}-\mu \right)\psi(\mathbf{r},t) - i\gamma(\mathbf{r})\left(\mathcal{L}_f-\mu\right)\psirt,
    \label{eq:GPEFull}
\end{equation}
where $\mathcal{L}_f \equiv \mathcal{L}-V(\mathbf{r})$.
The damping rate $\gamma(\mathbf{r})$ is ramped up smoothly from zero across the interface between the computational region and the fringe region to prevent reflections.  

\begin{figure*}[!t]
\includegraphics[width=\textwidth,trim= 0cm 0cm 0cm 0cm, clip=true]{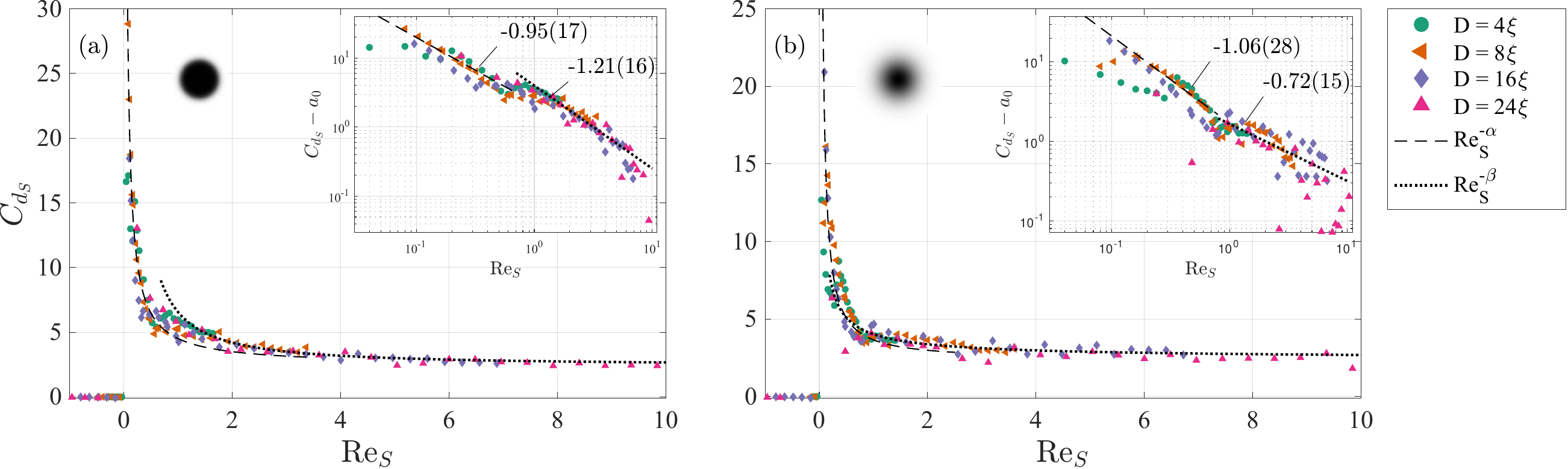}
\caption{Drag coefficient as a function of the Reynolds number for a (a) hard walled and (b) Gaussian obstacle. Data points of one colour/shape correspond to a constant obstacle size at different wake velocities. The collapse of data points of different obstacle sizes onto a single curve illustrate the presence of the dynamical similarity for superfluidic drag. The log-log insets plot $C_{d_S}-a_0 \sim C_{d_S} - C_{d_S}(\textrm{Re}_S\simeq\infty)$ as a function of $\textrm{Re}_S$ and reveal power laws for both obstacles corresponding $\alpha = 0.95(17)$ and $\beta = 1.21(16)$ for the hard walled obstacle, and $\alpha = 1.06(28)$ and $\beta = 0.72(15)$ for the Gaussian barrier. The outliers at small Reynolds number correspond to the quantum exclusive OD shedding regime and are not considered for the dynamical similarity.}
 \label{fig:2}
\end{figure*}

We first investigate the drag force for flow past a Gaussian obstacle, $ V_{G}(\mathbf{r}) = V_0 \exp \left( -r^2/\sigma^2 \right)$, as a function of the obstacle size. Gaussian obstacles are most readily available to experiments, and are implemented with a focused optical beam~\cite{Ketterle2001VortexLattice,Neely2010VortexDipole,Kwon2015CriticalVelocity}. The vortex shedding regimes were identified in Ref.~\cite{Sasaki2010} and are also shown in Figure~\ref{fig:fig1}(a): (I) the oblique dipole (OD), (II) the charge-2 von K\'arm\'an shedding (K2), and (III) irregular shedding. Their corresponding longitudinal force time series are plotted in (b). For superfluids, shedding becomes sufficiently irregular (III), corresponding to the transition to fully developed turbulence in the wake of the barrier, beyond $\textrm{Re}_S\sim0.7$~\cite{Reeves2015}. The drag force (see Figure~\ref{fig:fig1} (III,b)) time series resembles that of turbulent flow in this regime. 

Panels (c) and (d) of Figure~\ref{fig:fig1} show the drag force as a function of the superfluid velocity for a range of Gaussian obstacle sizes. In the OD regime, the drag force is dominated by the quantum vortex dipole momentum~\cite{pitaevskii2003bose} and increases linearly with the shedding frequency. In the K2 regime, two same-sign vortices shed from the top or the bottom of the barrier (Fig. \ref{fig:fig1}(II)). Shedding of a doublet occurs -- contrary to OD shedding -- out of phase between the top and the bottom of the barrier. The rapid shedding of a same-sign vortex pair results in a majority of the force being exerted into the transverse direction, resulting in a decreased linear gradient for the drag force, depicted in the shaded area labelled as ``K2" in Fig. \ref{fig:fig1}(c). The shaded area labelled as ``Irr" depicts the irregular shedding regime for the Gaussian barrier. Beyond $\textrm{Re}_\textrm{S}\sim0.7$, K2 shedding is unstable and shedding occurs irregularly and at a large frequency, causing clusters of vortices to form. The drag force increases at a faster rate in this regime.

The panels (c) and (d) also illustrate a qualitative difference in the shedding behaviour of small and large obstacles: (i) for increasingly larger obstacle sizes, the OD and K2 regime are unstable, thus absent, and irregular shedding starts immediately for $u>u_c$ (Fig. \ref{fig:fig1}d), and (ii) the drag force scales quadratically with the flow velocity, $F_d \sim u^2$. This quadratic scaling emerges due to the large number of vortex-vortex interactions around the barrier, which start to significantly affect the shedding and become the main component of the drag force. By contrast, for sufficiently small obstacles, the drag force is well described by the dipole momentum and provides the largest contribution to the drag, which is linear with $u$.

Next we probe the Reynolds law of dynamical similarity for superfluidic drag by repeating the numerical studies presented in Figure~\ref{fig:fig1} for different obstacle sizes and shapes and in each case extract the dimensionless drag coefficient $C_{d_S}$ as a function of the superfluid Reynolds number. The dimensionless drag coefficient expresses the momentum transfer of a fluid with an object of a particular geometry, independent of object size, fluid velocity and density, thus allowing us to extract the universal drag scaling law. The drag coefficient is given by 
$C_{d_S} = 2 F_d/\rho (u^2-u_c^2) D$, where we have modified the usual definition of the drag coefficient~\cite{hoerner1965fluid} to account for the absence of drag in a superfluid below the critical velocity $u_c$, as we describe in S2 of~\cite{supplementary}.

Figure \ref{fig:2} depicts the drag coefficient as a function of the Reynolds number for (a) a hard-wall circular barrier and (b) a Gaussian barrier. The collapse of the data points corresponding to different obstacle sizes onto a single curve is clear evidence of the dynamical similarity for superfluidic drag, with similar qualitative features as reported for those in classical hydrodynamics~\cite{roshko1954development,goldstein1938modern,wenDragTwodimensionalFlow2004}. To extract an empirical expression for the drag similitude of a particular geometry, we use the procedure used in classical hydrodynamics where the drag coefficient across different flow regimes is fit by a functional form~\cite{kaskas1970schwarmgeschwindigkeit}

\begin{equation}
    C_d = a_1\,\textrm{Re}^{-\alpha} + a_2\,\textrm{Re}^{-\beta}+a_0,
    \label{eq:EmpericalKaskasLaw}
\end{equation}
with $\alpha$ corresponding to the scaling in the low $\textrm{Re}$ limit, $\beta$ at an intermittent region and $a_0$ corresponding to the inertial asymptotic regime at $C_d(\textrm{Re}\rightarrow\infty)= a_0$. In three-dimensional systems, the persistence of the dissipation anomaly suggests $a_0 > 0$~\cite{Frisch1995Turbulence,babuin2014effective,galantucci2023dissipation}. For two-dimensional systems, the presence of a boundary provides a means for vorticity production, leading to finite drag even at $\textrm{Re}\rightarrow\infty$. This suggests $a_0>0$ for two-dimensional systems in the presence of an obstacle~\cite{clercxTwodimensionalTurbulenceSquare2001,drivasOnsagersConjecture2018,eyinkParadox2021}.

To prevent overparametrization, we fit $\alpha$ and $\beta$ in three steps: (i) $a_0$ is obtained by taking the mean of the last five datapoints. At this point, the curve is sufficiently flat with a sufficiently low standard error. (ii) To find $\alpha$ ($\beta$), $a_2$ ($a_1$) is set to zero. (iii) $\alpha$ is found at $\textrm{Re}_S<0.7$, $\beta$ is obtained for $\textrm{Re}_S>0.7$ on a log-log plot with $C_{d_S}-a_0$ as a function of $\textrm{Re}_S$. Outliers at low Reynolds numbers correspond to the superfluid unique OD vortex shedding regime and are not part of the similitude. Outliers at large Reynolds numbers are introduced artificially through subtracting $a_0$ and not considered in fitting of the power laws. We included the log-log plots in which $a_0$ is not subtracted in S3 of~\cite{supplementary}.

The main source of error in our fits is the confidence in estimation of the asymptotic constant, $a_0$, which we extract from our numerical simulations. This is due to the limited range of computationally accessible Reynolds numbers; a large range is required to demonstrate the fully inertial regime at which $a_0$ may be obtained with higher confidence.

A number of striking parallels exist between the quantum and classical dynamic similarity. At low $\textrm{Re}$, the classical sphere and cylinder obstacles' drag coefficient follow $C_d \approx \textrm{Re}^{-1}$, i.e. $\alpha = -1$.~\cite{Stokes1851,cylinderCorrection,lamb1911xv}, similar to the superfluidic hard wall and Gaussian obstacles, with $\alpha = 0.95(17)$ and $1.06(28)$, respectively (see inset of Fig. \ref{fig:2}). The underlying mechanisms causing this power law are vastly different, with the classical $\alpha$ originating from time-reversible Stokes flow. In contrast, below the speed of sound, ($u\ll c$), superfluidic drag originates almost exclusively from time-irreversible vortex nucleation dominated by the quantum dipole force~\cite{Frisch1992} which describes the OD and K2 regime well. The power laws at larger $\textrm{Re}_S$ correspond to $\beta = 1.21(16)$ and $0.72(15)$ for the hard wall and Gaussian respectively, and are the result of more complex vortex dynamics occurring in the irregular shedding regime. We provide a more complete picture of how the power laws $\alpha$ and $\beta$ correspond to the different shedding regimes in S4 of~\cite{supplementary}. The magnitude of $\beta$ with respect to $\alpha$ suggests vortex shedding is more efficient in the irregular regime than in the K2 regime for the hard walled obstacle. In contrast, irregular shedding is less efficient for the Gaussian. This explanation of
the behavior is supported through analysis of the Strouhal number~\cite{Reeves2015}, as described in S4 of~\cite{supplementary}.

\begin{table}[b!]
%\resizebox{\columnwidth}{!}{%
\begin{tabular}{@{}cccc@{}}
\toprule[1pt]
           &  & Classical ($\overline{C_d}$) & ~Quantum ($\overline{C_{d_S}}$) \\ \midrule[1pt]
Flat plate & {\includegraphics[width=0.04\textwidth,trim= 0cm 0.5cm 0cm 0cm]{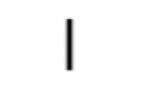}} & 2.3~\cite{liuDragForceAccelerating2024}            & 2.14(5)             \\
Wedge      & {\includegraphics[width=0.04\textwidth,trim= 0cm 0.5cm 0cm 0cm]{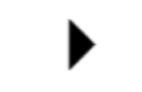}} & 1.7~\cite{mahatoDirectSimulationSound2018}            & 1.93(6)             \\
Circle     & {\includegraphics[width=0.04\textwidth,trim= 0cm 2.5cm 0cm 0cm]{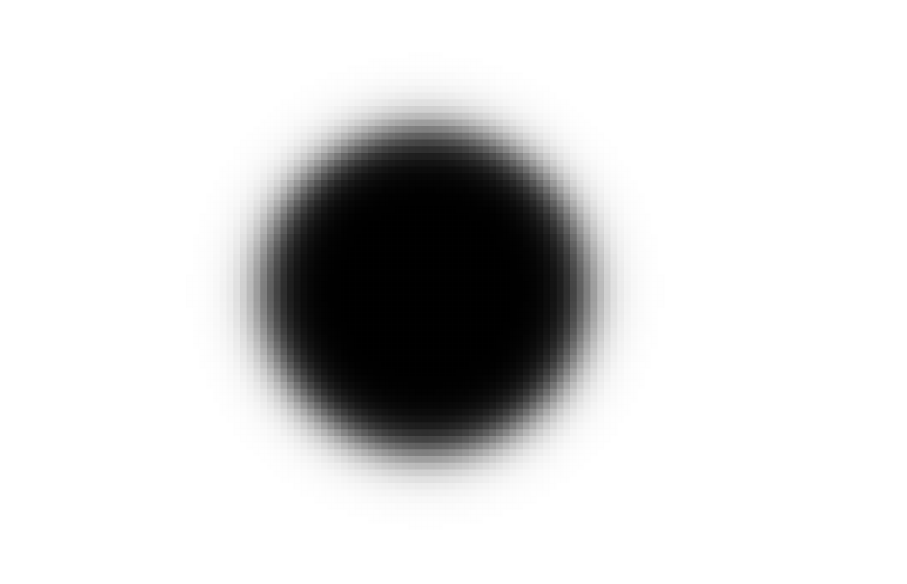}} & 1.4~\cite{dengDragForceCircular2022}            & 2.40(6)             \\
Gaussian   & {\includegraphics[width=0.04\textwidth,trim= 0cm 0.3cm 0cm 0cm]{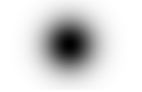}} & Not applicable             & 2.37(16)            \\
Airfoil    & {\includegraphics[width=0.04\textwidth,trim= 0cm 0.5cm 0cm 0cm]{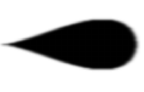}} & 0.09~\cite{nastechnicalNACA0012airfoil}            & 1.57(8)             \\ \bottomrule[1pt]
\end{tabular}%
%}
\caption{Drag coefficients at the large Reynolds number asymptotic limit, $\overline{C_d}$, in two-dimensional classical and quantum systems. The Gaussian obstacle has no trivial classical analog. Classical drag coefficients obtained from \cite{liuDragForceAccelerating2024,mahatoDirectSimulationSound2018,dengDragForceCircular2022,nastechnicalNACA0012airfoil}.}
\label{tab:dragCoeffs}
\end{table}

In addition to the hard-wall circular and Gaussian obstacle, we confirm the drag similitude and investigate the asymptotic $a_0$ value for non-trivial obstacle shapes. In classical hydrodynamics, $\overline{C_d} = a_0$ is the number associated with an object's `constant' drag coefficient and quantifies how aerodynamic an object's shape is. Table~\ref{tab:dragCoeffs} summarizes the constant drag coefficient for a flat plate, wedge, and airfoil-shaped obstacle and compares it to the classical values. All obstacles are axisymmetric and approached under a zero degree angle of attack to prevent lift-induced drag. The drag coefficients have been extracted using the same procedure as those of the Gaussian and hard-wall obstacles. We remark on the stark difference between the classical and superfluid airfoil. The drag coefficient is significantly lower than for the other superfluid obstacles due to delayed flow separation (also identified in Ref. \cite{musserAirfoilLift2019}), but happens to be much less efficient than for the classical airfoil. Despite the marginally larger drag coefficient, the airfoil is significantly more streamlined than the other obstacles: The critical velocity of the obstacle is significantly higher than that of bluff bodies of the same diameter (see Table S1 of ~\cite{supplementary}). This suggests drag only emerges at much larger velocities for superfluid streamlined bodies than bluff bodies. We provide more information in S5 of~\cite{supplementary}, along with the drag coefficient curves for the various other obstacles.

Next, we outline how the universality of the drag coefficient can be verified experimentally. Here, we must consider the limited resolution with which we are able to probe the superfluid wake. This means that the procedure used in our numerical simulations, based on the Ehrenfest relation, cannot be directly applied experimentally. Instead we note that in an otherwise unperturbed flow, drag force may equivalently be obtained by considering the total momentum flow in a set volume around an obstacle. 

The first step is to cast the system into a control volume, as described in classical hydrodynamics~\cite{cengel2018FluidBook,NANGIA2017437ControlVolume}, using the Leibniz-Reynolds transport theorem~\cite{ReynoldsMechanicalAndPhysicalSubjects1903}:

\begin{align}
    \frac{\partial(m \mathbf{u})_{\textrm{sys}}}{\partial t} = \frac{\partial}{\partial t} \int_{\textrm{CV}} \rho \mathbf{u}\,dV + \int_{\textrm{CS}} \rho \mathbf{u} \left(\mathbf{u} \cdot \mathbf{n}\right)\,dA, 
    \label{eq:CV}
\end{align}
where the left hand side is the momentum flux of a closed system, CV is a control volume around the obstacle, and CS is the perimeter of the control volume. Here $\rho$ and $\mathbf{u}$ represent the density and velocity of the fluid inside the control volume, respectively, and $\mathbf{n}$ the normal vector pointing out of the control surface. The first term on the right-hand side is the change of total momentum inside the control volume in time, which vanishes in steady state with constant inflow velocity and density, and the second term on the right-hand side is the net flow of momentum through the control volume boundary.

As the next step, we rewrite Eq.~\eqref{eq:CV} in the quantum hydrodynamics approximation by replacing the density $\rho \to m \psi^*\psi$, and $\rho\mathbf{u}\to \mathbf{J} = \frac{-i \hbar}{2}\left(\psi^*\nabla\psi - \psi\nabla\psi^*\right)$~\cite{winiecki2000vortex}. For a two-dimensional system the resulting expression for the net force acting on the control area, or the momentum flux, is given by
\small
\begin{equation}
\frac{\partial}{\partial t} \varoiint_\Omega \mathbf{J}\,d\Omega + \oint_\Gamma \mathbf{J}(\mathbf{u}\cdot\mathbf{n})\,dl = \oint_\Gamma T_{jk}\, d\mathbf{l} - \iint_\Omega \rho \partial_k \left(\frac{V}{m}\right)\,d\Omega, 
\label{eq:CVQuantum} 
\end{equation}
\normalsize
where $\Omega$ is the two-dimensional control area, $\Gamma$ is the surface encompassing the control area, and $T_{jk}$ is the momentum flux density tensor~\cite{Frisch1992}. The last term is the Ehrenfest relation in the hydrodynamic limit, with $\partial_k$ the spatial derivative in $k$ and $V$ the potential and $m$ the mass~\cite{winiecki2000vortex}. 

In the steady state and for sufficiently smooth obstacles, where the above equation can be further simplified to
$$
 \oint_\Gamma \mathbf{J}(\mathbf{u}\cdot\mathbf{n})\,dl = - \iint_\Omega \rho \partial_k \left(\frac{V}{m}\right)\,d\Omega,
$$
the force in direction $k$ may be evaluated using the Ehrenfest relation, or by considering the momentum current density flowing through a control surface around an obstacle. This control volume method introduced here has great experimental utility as the largest contributions to the Ehrenfest relation occur within a number of healing lengths around the obstacle, requiring large experimental resolution to obtain the drag force accurately.  

\begin{figure}[]
  \centering
 \includegraphics[width=1\columnwidth]{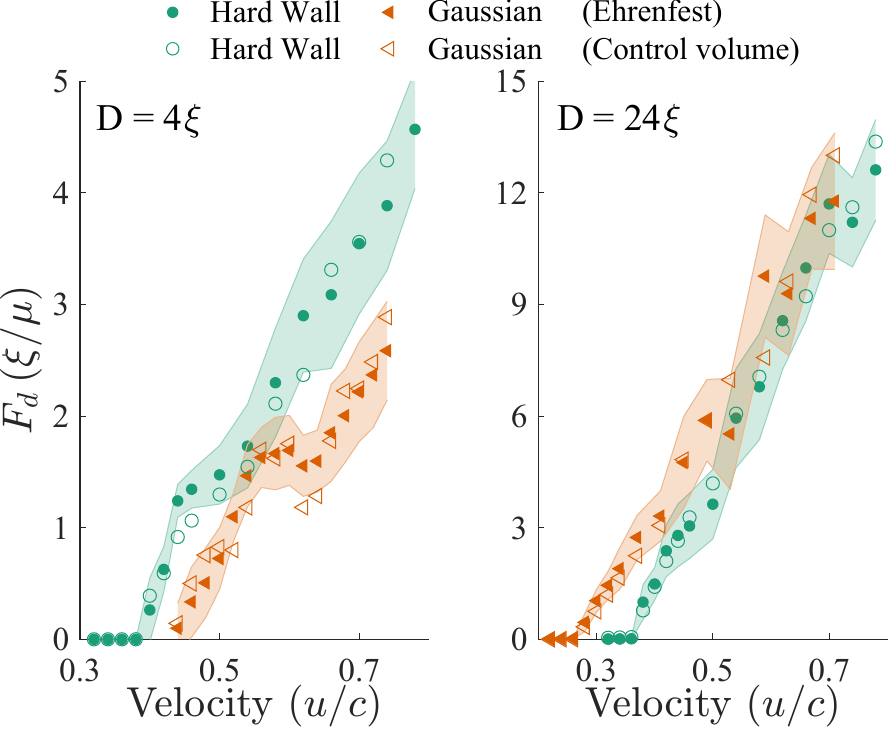}
 \caption{Drag force for a hard wall (circle) and Gaussian (triangle) obstacle, obtained from the longitudinal force time series using the Ehrenfest relation (filled datapoints) and control volume method (open dots). Shaded areas represent the error on the Ehrenfest method.}
 \label{fig:4}
\end{figure}

We have analyzed the data from Fig. \ref{fig:fig1}(c,d) using the two methods, and the results are plotted in Fig. \ref{fig:4}. We limit the downstream boundary to be no further than $L_y/2$ away from the obstacle to prevent contributions from edge effects in our finite-sized grid.  We compare the method for small (left) and large (right) obstacles, for both the hard wall (green dots) and Gaussian (orange triangles) barriers. The filled data points correspond to the Ehrenfest method, with corresponding shaded areas defining the error. Due to long time correlations, the error is equivalent to the standard deviation of the force time series (depicted in Fig. \ref{fig:fig1}(b)). Open data points are obtained through the control volume method. Across the range of all our independent parameters, the control volume method is in good agreement with the Ehrenfest method, showing the effectiveness of this method. This method will open up new in situ techniques for measuring superfluid drag in Bose-Einstein condensates: Experimental verification is realisable for superfluids with access to the velocity field, such as atomic Bose-Einstein condensate experiments utilising Bragg spectroscopy~\cite{navon2016BraggMomentum,seo2017BraggVelocity}.

In conclusion, we have confirmed the existence of the Reynolds law of similarity for superfluidic drag. Existence of the drag similitude provides further evidence of the utility of the superfluid Reynolds number. Our results, and the earlier demonstrated universal behaviour for the Strouhal number~\cite{Reeves2015}, suggest \textit{all} relevant dimensionless quantities are expressible as a universal function of the $\textrm{Re}_S$ due to scale invariance. In addition, we investigated the drag coefficient of various bluff and streamlined obstacles and found significant differences with a significantly higher drag coefficient for a streamlined airfoil than its classical equivalent. Finally, we provide an alternative method of obtaining the drag force in a wake which opens up the ability for experimental measurements of superfluidic drag. Future work at larger Reynolds numbers, realised at much larger computational grids, can shed new light on constant drag coefficients and on dynamical behaviour in the fully inertial regime, where $\textrm{Re}_S \sim \textrm{Re}$.

\textit{Acknowledgements}---We thank G. Gauthier for the valuable discussions. This research was supported by the Australian Research Council (ARC) Centre of Excellence for Engineered Quantum Systems (EQUS Grant No. CE170100009), and ARC Future Fellowship FT190100306. M.T.M.C. acknowledges the support of an Australian Government Research and Training Program Scholarship. A.S.N. is supported by the Dutch Research Council (NWO/OCW), as part of the Quantum Software Consortium programme (project number 024.003.037) and Quantum Delta NL (project number NGF.1582.22.030),  T.W.N. acknowledges the support of Australian Research Council Future Fellowship No. FT190100306. M.T.R. acknowledges the support of an Australian Research Council Discovery Early Career Research Award DE220101548. Computing support was provided by the Getafix cluster at the University of Queensland.

\textit{Data availability}---The data generated for this Letter are
available at~\cite{data} and include all force time series, processing code, and processed data from which all relevant parameters have been obtained. Because of the large amount of storage required, the raw wave functions are not available. Code used to generate the wave functions is
available upon request.

%\bibliography{main}{} 

\begin{thebibliography}{57}
\bibitem [{dis()}]{dissipationAnomalyNote}%
  \BibitemOpen
  \href@noop {} {}\bibinfo {note} {We note that while the dissipation anomaly has been both observed in classical and zero-temperature quantum turbulence, classical dissipation of incompressible kinetic energy occurs due to viscous effects. In contrast, the anomaly can persist in the zero-temperature case due to the compressibility of the superfluid; the phonon bath provides a small scale sink for the dissipation of the (incompressible) energy in the vorticity.}\BibitemShut {Stop}%
\bibitem [{SFR()}]{SFReynolds}%
  \BibitemOpen
  \href@noop {} {}\bibinfo {note} {Note, we choose $\kappa = \hbar/m$, consistent with our choice of natural units, instead of the conventional $\kappa = h/m$. This results in a transition to turbulence at $\textrm{Re}_S \sim 1$}\BibitemShut {NoStop}%
\end{thebibliography}

%

\clearpage

\foreach \x in {1,...,5}
{%
\clearpage
\includepdf[pages={\x}]{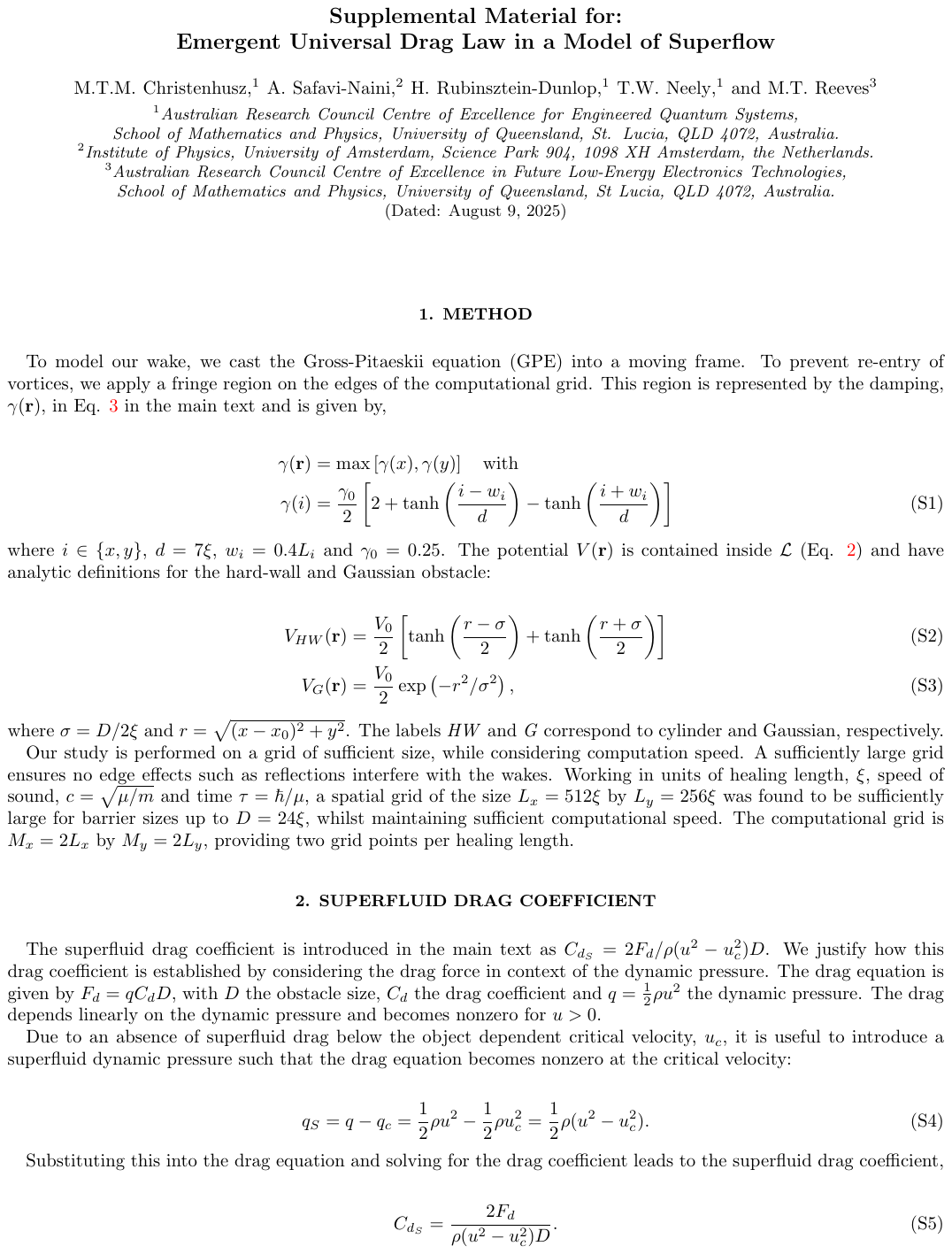} 
}

\end{document}